\documentclass{llncs}

\usepackage{xcolor}
\usepackage{graphicx}
\usepackage{multirow}

\newcommand{\comment}[1]{\iffalse #1 \fi}
\newcommand{\isau}{{\sf isabelle-users}}
\newcommand{\coqc}{{\sf coq-club}}
\pdfoutput=1

\begin{document}
\raggedbottom

%
%
\pagestyle{headings}  

%
%
\title{Social Network Processes in the Isabelle and Coq Theorem Proving Communities
}
\titlerunning{Interactions in the Isabelle and Coq communities}  
%
\author{Jacques Fleuriot \and Steven Obua \and Phil Scott}
\authorrunning{Jacques Fleuriot et al.} 

\institute{Informatics Forum, University of Edinburgh \\ 
10 Crichton Street, Edinburgh, EH8 9AB, UK\\
\texttt{http://www.proofpeer.net}
}

\maketitle              

\begin{abstract}
 
We identify the main actors in the Isabelle and Coq communities and describe how they affect and influence their peers. 
This work explores selected foundations of social networking analysis that we expect to be useful in the context of the ProofPeer project, which is developing a new model for interactive theorem proving based on collaboration and social interactions.
 
\end{abstract}
\section{Introduction}

This paper will mention people in the theorem proving world. It may even refer to you. So, let us start with a caveat stating that no personal criticism is intended in cases where actual events and behaviours are described. These merely serve to provide concrete illustrations of some of the social processes in the Isabelle and Coq communities that our analysis has unveiled. 

Interactions and knowledge exchange within the major theorem proving communities occur mainly via mailing lists. In the current work, we describe our exploration of the email exchanges on the Isabelle-users and Coq Club mailing lists, which constitute the  two largest and most active groups of interactive theorem proving users. We study connectedness  at the network structure level by looking at links between individuals and also examine how various individuals' actions can have consequences for other members of the network.

The paper is organised as follows: in the next section, we give an overview of our datasets and how they were obtained. In Section 3, we  study the two communities in order to gain an understanding from a foundational but effective social network analysis standpoint. In Section 4, we look at the notion of sub-communities and give evidence of their existence in our networks. Finally, in Section 5, we offer a few conclusions and briefly place this work within its broader ProofPeer context. 

\section{Raw Data and Extraction Process}

Our raw data from the publicly-available Isabelle (\isau) and Coq (\coqc) mailing lists can be summarised as follows:
\begin{itemize}
\item \isau:  11 818 messages, 112 MB, August 2008 to August 2015.
\item \coqc: 15 886 messages, 107 MB, August 2004 to September 2015.
\end{itemize}

	These messages were gathered by downloading the mailing list data for the periods shown.\footnote{We thank Hugo Herbelin for providing the \coqc{}  messages  in a readable format.} We excluded messages predating August 2008 for \isau{} due to various issues that caused unreliable parsing of the archived data prior to this date. All data was converted to the well-known MBox format\footnote{https://tools.ietf.org/html/rfc4155}, a task that required some care since the two communities use different mailing list managers (Mailman for \isau{} and Sympa for \coqc) and, moreover, the mail clients of some of the users messed up the message headers in some cases.

Once the archives are in MBox format, we extract, for each message: its unique identifier, its sender, its timestamp, and the subject. Other extracted fields shall be described later as appropriate. It is worth noting that a fair amount of time is needed to systematically clean the extracted data. Among other things, we must deal with the different character encodings (e.g. utf-8 vs. ascii vs. iso8859-1), convert dates in order to take into account time zones (and work out the message order in some cases), and remove MIME attachments (e.g. caused by users attaching screenshots on \isau{}) and HTML tags.
\begin{figure}
\begin{center}
\includegraphics[width=1.0\textwidth]{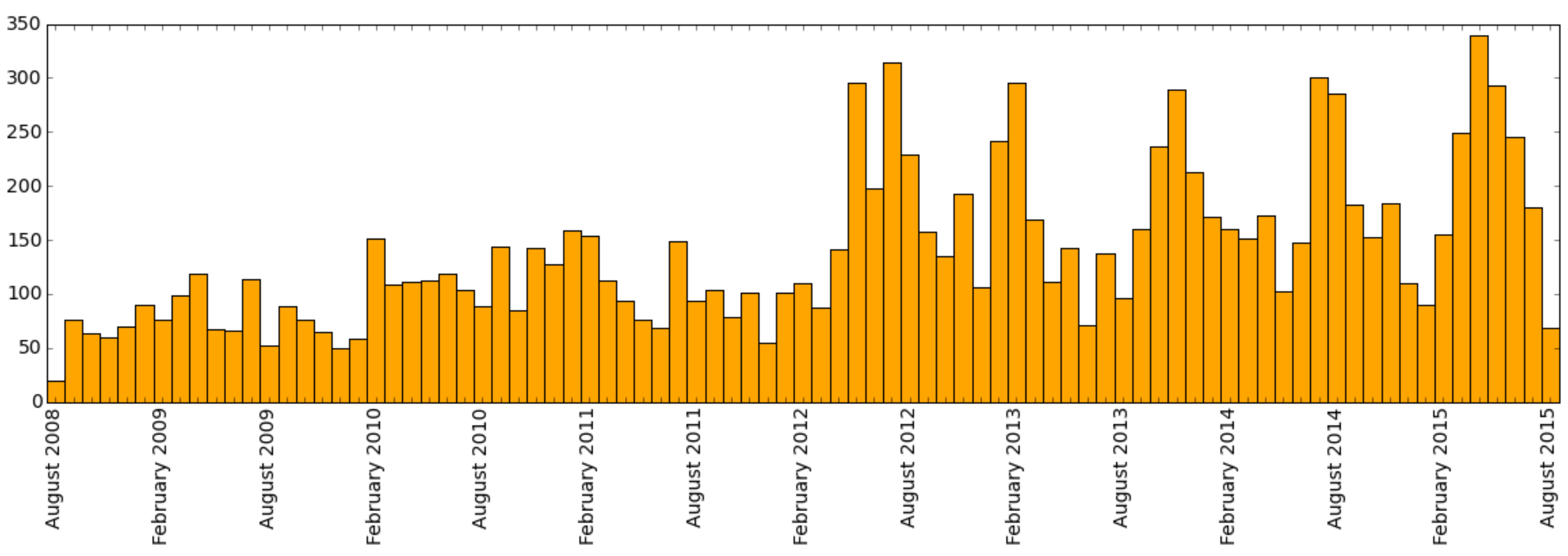}
\end{center}
\caption{An overview of the number of Isabelle messages exchanged per month, with an average of 143 messages/month and a peak of 339 messages in April 2015.}
\label{fig:isabelle-messages-overview}
\end{figure}

\begin{figure}
\begin{center}
\includegraphics[width=1.0\textwidth]{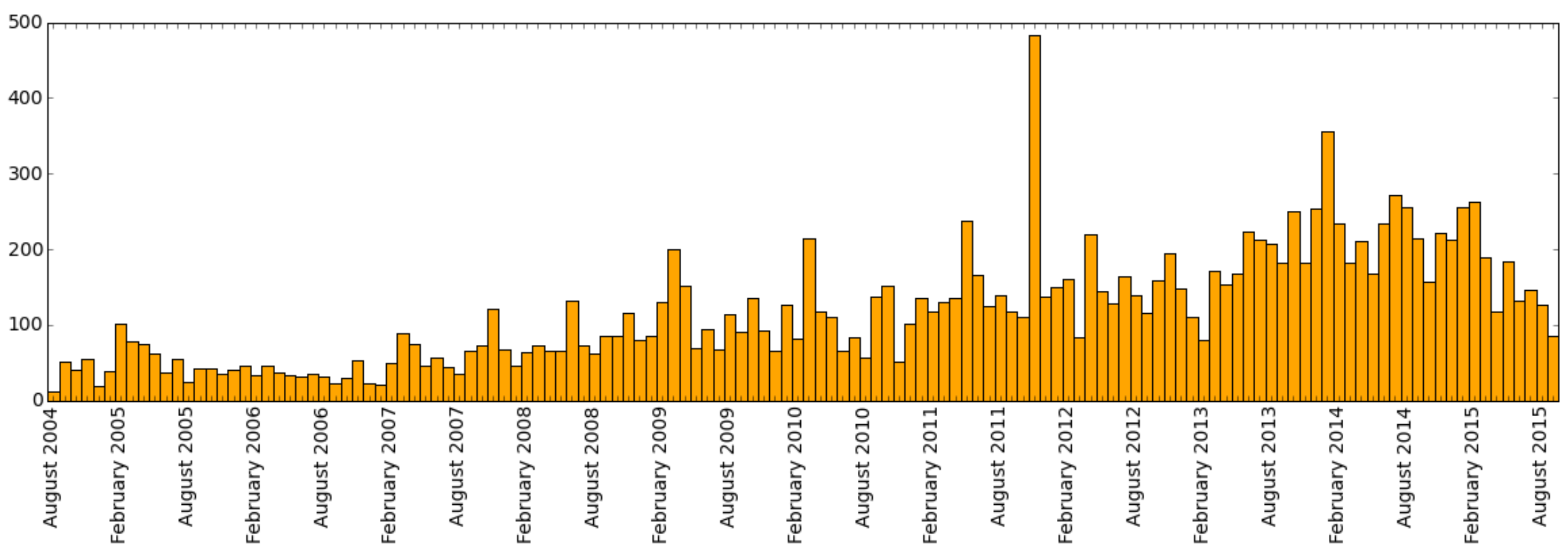}
\end{center}
\caption{An overview of the number of Coq messages exchanged per month, with an average of 159 messages/month and a peak of 483 messages in November 2011.}
\label{fig:coq-messages-overview}
\end{figure}

Figures \ref{fig:isabelle-messages-overview} and \ref{fig:coq-messages-overview} give an idea of the distribution of messages over the time periods mentioned above. The April 2015 peak in the number of \isau{} messages coincides with Isabelle2015-RC0 being made available for testing, but we identified no particular event to explain the peak of 483 messages in November 2011 (the longest thread for that month was about set theory and consisted of 26 messages). Overall, it is easy to spot that both \isau{} and \coqc{} have been growing over the past few years, which serves the current purpose of this work.

\subsection{Brief Overview of Processed Data}

In this section, we give a summary of the data once it has been processed. In particular, we reconstruct the message threads using an unpublished but well-known algorithm called \emph{jwz threading} that is generally considered a viable approach to parsing message threads\footnote{The algorithm can be found at http://www.jwz.org/doc/threading.html}. Once the reconstruction is done, we only consider threads of \emph{2 messages or more}. We thus rule out most posts related to things such as calls for papers and job openings, as well as posts that receive no replies, these not being relevant to the current work. Besides, a cursory analysis reveals that messages requiring an answer but receiving none are relatively rare on both lists. We end up with the following datasets for our two mail corpora:

\begin{itemize}
\item \isau:  2247 threads, 10 388 messages, 600 authors.
\item \coqc: 2714 threads, 12 947 messages, 1244 authors.
\end{itemize}

We note for the above that a substantial amount of effort was spent on working out the authors of the messages. This relied on both names and email addresses, applying various disambiguation approaches including the use of Levenshtein distance \cite{levenshtein1966bcc}. Much of it could be done automatically but a fair amount of time was also spent manually inspecting and fixing the data using our knowledge of the Isabelle and Coq communities.  

All the analysis that follows will be based on information extracted from these two datasets, which we convert into JSON format \cite{json-ecma13} and store in a MongoDB database \cite{Chodorow:2010:MDG:1941134} since this makes retrieval and manipulation particularly convenient. 

\section{Basic Social Network Analysis}

In this section, we start the actual  analysis of social processes on the \isau{} and \coqc{} lists. Our results will be about the members of the communities but also about what they discuss and the connections that are formed around particular topics.  

\subsection{Possible Communication Models}

A variety of  choices are possible when it comes to building the communication models for a mailing list. For instance, one may go for a simple Q\&A type one where the author of the opening post is connected by an edge to all the authors who reply in the thread. Another model, (e.g.\ the one adopted by Bohn et al.\ \cite{bohn-r-ml11}) connects anyone who replies to an email to all the other authors in the  thread before them, with the assumption that a respondent is aware of all the previous emails. However, we believe that neither of these two models fully captures the dynamics of the actual interactions. For instance, it is pretty clear that respondents often do not read every message in a thread before replying. 

Instead, we make use of all the information that is available from our thread construction to connect authors in a more discriminating way. In particular, we use the fact that threads are non-linear (i.e. replies often contain explicit information about the intended recipient) and use this to create a directed edge from individual $A$ to individual $B$ if $A$ sent a message to $B$. In cases where the intended recipient is not explicit, we assume that the person is replying to the individual who made the opening post. We construct our directed graphs using NetworkX \cite{hagberg-2008-exploring}, a Python software package for the study of the structure, dynamics and functions of complex networks.

\subsection{A Macro-level View: Reciprocity}
\label{section:reciprocity}
In a directed network, \emph{reciprocity} refers to the notion that two actors are mutually connected to each other~\cite{wasserman1994social}. With our directed data, there are four possible connection relationships: $A$ and $B$ never communicated, $A$ sent a message to $B$, $B$ sent a message to $A$ or $A$ and $B$ have sent messages to each other.    

More concretely, the standard notion of reciprocity (SR) for a directed graph can be defined as the ratio of the number of edges pointing in both directions to the total number of edges in the graph. However, this measure cannot distinguish the relative difference in reciprocity when compared with a purely random network with the same number of nodes and edges \cite{costa-char-2007}.  So, in addition, we use the Garlaschelli and Loffredo notion of edge-reciprocity (GLR) \cite{PhysRevLett.93.268701} that aims to deal with this problem (we refer the interested reader to the paper for  the mathematical details) in order to gain an idea of how social our theorem proving communities are, compared to each other and to a few well-known networks, irrespective of relative size.  Our analysis reveals that under both  measures (excluding self-loops explicitly in the case of SR) the networks exhibit reciprocity: 
\begin{itemize}
\item \isau{}: SR:  0.474; GLR:  0.469 
\item \coqc{}: SR: 0.367; GLR: 0.365 
\end{itemize}
Based on these results, \isau{} have more actors who are willing to engage in both asking and replying to messages than \coqc{}. This appears to indicate that the \isau{} community is more ``inclusive", with fewer theorem proving ``gurus" than the \coqc{} community. A simple, though unverified, reason for this may lie in the fact that mastering Coq's dependently-typed setting in order to reply to messages demands a lot of expertise compared to Isabelle (where many messages are about higher-order logic). For a broader comparison (even though the connections between vertices of the graph have different meanings), we note that Flickr (in 2009) had a reciprocity of 0.68 \cite{Cha:2009:MAI:1526709.1526806}, YouTube  had one of  0.79 (in 2007) \cite {Mislove:2007:MAO:1298306.1298311} and Twitter, with its followers-based model,  had a reciprocity of only 0.22  (in 2010) \cite{kwak:2010:TSN:1772690.1772751}.  Viewed in relation to these well-known communities, this indicates that our mailing lists are relatively social when it comes to reciprocal relationships.

\subsection{Power Models}

We now investigate the topology and dynamics of our two communities by identifying some its most important actors via \emph{centrality} measures; i.e.\  how `central' particular nodes are to the network based on different metrics.  We start with degree centrality, which is the number of ties a node has. This is a simple yet powerful concept that can be used to characterise notions such as prestige and power.

\subsubsection{Degree Centrality}

\begin{figure}
\begin{center}
\includegraphics[width=1.0\textwidth]{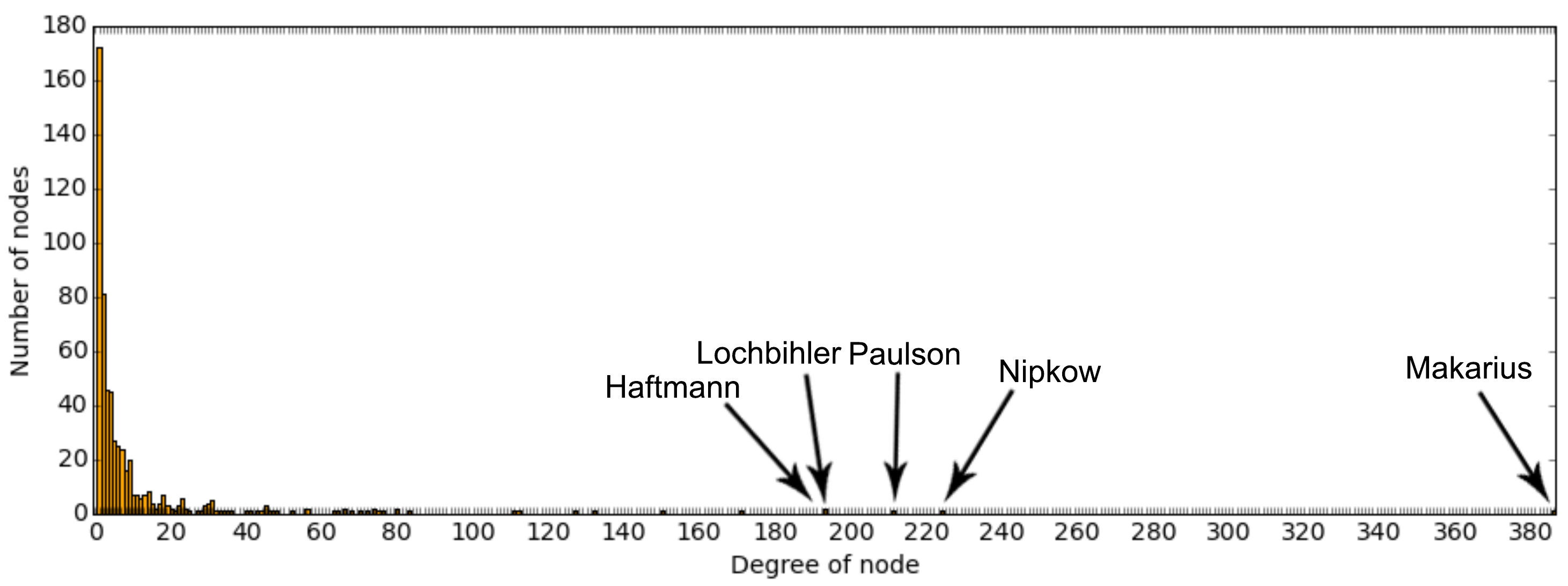}
\end{center}
\caption{Overall degree connectedness for \isau{}.}
\label{fig:isa-deg-connect}
\end{figure}

\begin{figure}
\begin{center}
\includegraphics[width=1.0\textwidth]{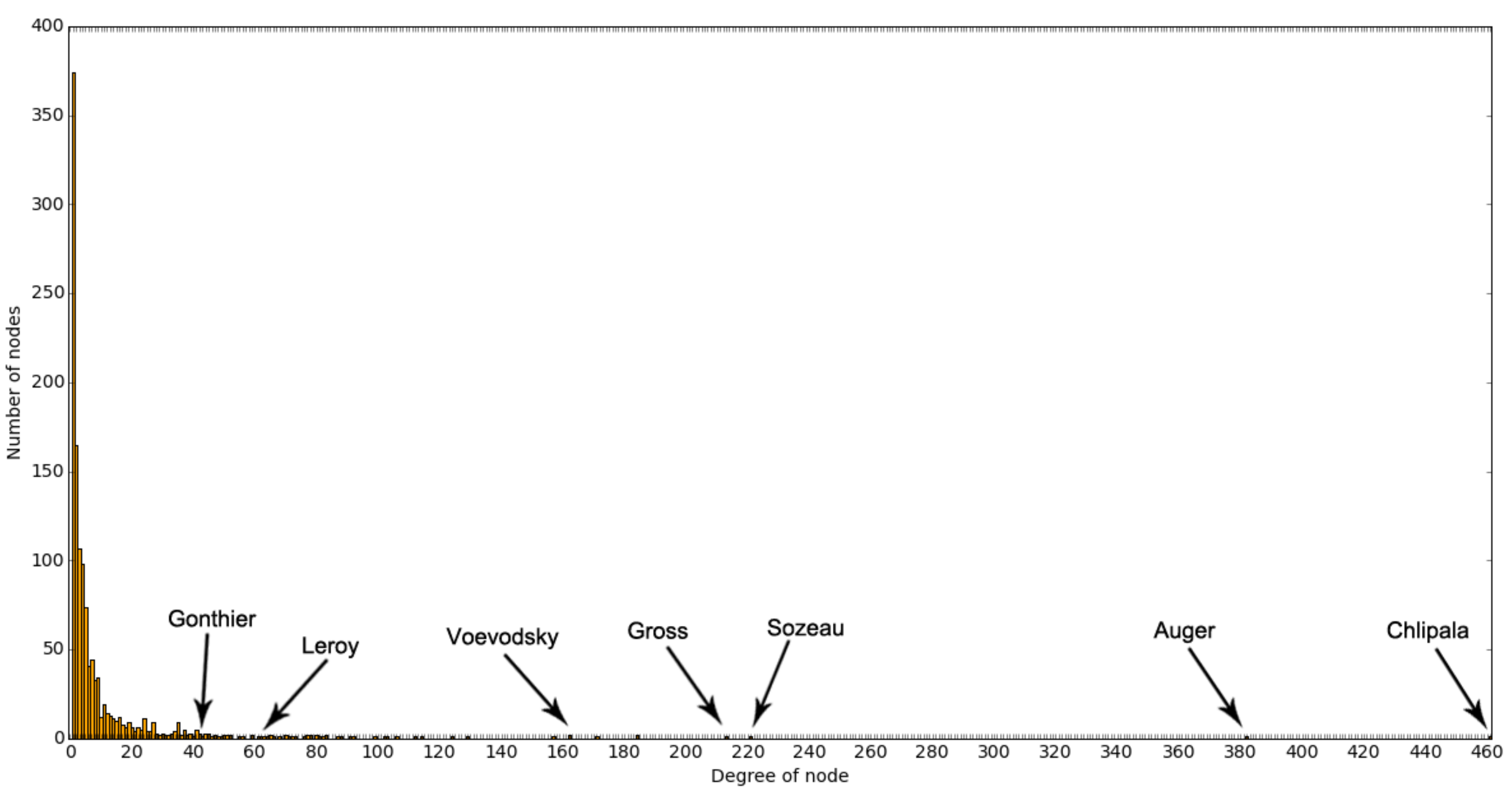}
\end{center}
\caption{Overall degree connectedness for \coqc{}.}
\label{fig:coq-deg-connect}
\end{figure}

The histograms in Figures \ref{fig:isa-deg-connect} and \ref{fig:coq-deg-connect} give an overview of the overall level of interaction via the sum of in- and out-degree connectedness for  \isau{} and \coqc{}. The communication in each case displays a typical heavy-tailed characteristic, with most people interacting with just one (or two) individuals directly and a few who interact with many.  

\comment{ REMOVE?
In order to gain a better understanding of the social in-degree and out-degree of the two communities, we decided to identify best-fit distribution for the degree-connectedness data. We found that unlike many studies on social networks, in which the in- and out-degrees tend to follow a power-law distribution \cite{Mislove:2007:MAO:1298306.1298311}, the degrees in our case seem best best captured by a discrete lognormal distribution.  Should we say more here? If so, add graphs for fits?}

However, with our directed data, we can go one step further and distinguish centrality based on in-degree from that based on out-degree.\footnote{We note, in passing, that there are modifications of this simple idea (e.g.\ that proposed by Bonacich \cite{power-bonacich87}) that we also looked at but will not discuss as they do not add much to the current analysis.} In general, an individual with a high in-degree is usually considered to play a prominent role in the community and/or have a high prestige. In our case, this can be viewed as indicating someone who sends interesting or important posts and thus gets many replies. Note that this does not necessarily mean that they send many messages (see below).  Individuals with a high out-degree are those who exchange messages with many others and/or are opinionated and like to make many others aware of their views. Such individuals are often viewed as being very active and influential in the network. With these characterisations in mind, we highlight the members of the two communities with the highest in- and out-degree centralities in Figures~\ref{fig:indegrees-actors} and \ref{fig:outdegrees-actors} and based on these make a few concrete remarks.

\begin{figure}
\parbox{.45\linewidth}{
\centering
\begin{tabular}{|@{\hskip 0.2cm}c@{\hskip 0.2cm}|@{\hskip 0.2cm}l@{\hskip 0.3cm}|@{\hskip 0.2cm}l@{\hskip 0.2cm}|}
\hline
& \isau{} & \coqc{} \\ \hline
1 &  Makarius & Chlipala \\ 
\hline
2 &  Lochbihler &  Auger \\ 
\hline
3 &  Nipkow &  Voevodsky \\ 
\hline
4 &  Huffman & Gross \\
\hline
5 &  Paulson & Sozeau \\
\hline
6 &   Haftmann & Vries \\ 
\hline
7 &   Sternagel & Spiwack   \\ 
\hline
8 &  Lammich&  Casteran    \\ 
\hline
9 &   Blanchette & Schepler   \\
\hline
10 & Noschinski  & Braibant   \\
\hline
\end{tabular}
\textnormal{\caption{In-degree centrality ranking \mbox{}\mbox{}\qquad  \label{fig:indegrees-actors}}}
}
\hfill
\parbox{.55\linewidth}{
\centering
\begin{tabular}{|@{\hskip 0.1cm}c@{\hskip 0.1cm} | @{\hskip 0.2cm}l@{\hskip 0.2cm} |@{\hskip 0.1cm}r@{\hskip 0.1cm}|@{\hskip 0.2cm}l@{\hskip 0.2cm}|@{\hskip 0.1cm}r@{\hskip 0.1cm}|}
\hline
& \isau{} &{\sf msgs} & \coqc{} & {\sf msgs} \\ \hline
1 &  Makarius & 1775 &  Chlipala & 789 \\ 
\hline
2 &  Nipkow & 589  &  Auger & 637 \\ 
\hline
3 &  Paulson & 401  &  Sozeau & 288 \\ 
\hline
4 &  Haftmann & 494 & Courtieu & 222 \\
\hline
5 &  Lochbihler & 535 & Gross & 450 \\
\hline
6 &   Huffman & 269 & Spiwack & 238\\ 
\hline
7 &  Sternagel & 351 & Herbelin   & 201\\ 
\hline
8 &  Noschinski & 331 &  Schepler & 234 \\ 
\hline
9 &  Blanchette & 215 &  Casteran & 206 \\
\hline
10 &  Krauss & 136 &  Blanqui & 159 \\
\hline
\end{tabular}
\textnormal{\caption{Out-degree centrality ranking with number of messages sent by each actor  \label{fig:outdegrees-actors}}}
}
\end{figure}

It is interesting to note that the Fields medallist, Vladimir Voevodsky, is ranked so highly despite sending only 190  out of the 12947 messages in our dataset; i.e.\ around 1.5\%. This can be contrasted with Chlipala, for instance, who is responsible for around 6.1\% of the messages sent on \coqc{}. When it comes to \isau{}, Makarius, who contributes around 17.1\% of the messages in our dataset, is the most prominent actor and also the one who interacts with most people. Based on an examination of the messages, Makarius is in all likelihood one of the most openly opinionated members of the community as well. 
Another point worth noting is that Gottfried Barrow (which we believe to be a pseudonym), who accounts for some 3.9\% of the Isabelle messages, is only ranked $18^{th}$ by in-degree and $22^{nd}$ by out-degree. These simple measures seem to confirm the previously anecdotal view that this user only sends mildly interesting messages that do not elicit much engagement from the \isau{} community. Overall, in our datasets, the top 10 actors per out-degree centrality account for around $49.1\%$ of the messages on \isau{}  (although this drops to around 32\% if we ignore Makarius) but only for around $26.4\%$ of the messages on \coqc{}.

\subsubsection{Betweenness Centrality}
\label{subsubsection:bc}
We now consider the notion of betweenness centrality (BC), which from a graph theoretic standpoint, simply measures the number of times a node acts as a bridge along the shortest path between two other nodes but, from a social network analysis standpoint, identifies nodes as `gatekeepers', where actors with a high betweenness centrality can \emph{control} or even \emph{disrupt} communications or trust relationships between various end points. Formally, the betweenness centrality $C_B(v)$ of a node $v$ is given by \cite{bet-centrality-freeman77}:
$$
C_B(v) = \sum_{s, t\in V \setminus \{v\}} \frac{\sigma_{st}(v)}{\sigma_{st}}
$$
Here $V\setminus\{v\}$ is the set of nodes except $v$, $\sigma_{st}$ is the number of shortest paths from $s$ to $t$ and $\sigma_{st}(v)$ is the number of those paths passing through $v$. 

Within the context of our mailing lists, we generally see that once an actor with a high betweenness centrality enters the conversation, they potentially get more replies addressed directly to them, making them important in-betweens in the ensuing discussion. We consider two examples from \isau{}. In particular, a  thread from November 2012 about ``Length of Proofs" started by Doll (ranked $30^{th}$  when considering BC)  involves only 1 message, out of 22, being sent explicitly to him while 6 are sent as direct replies to Paulson and 4 directly  to Makarius once they enter the conversation.

\noindent As an example of a disruption that demonstrates a gatekeeping role in action, we have the message sent by Schmalz  about ``parse translations and lexical matters'' from January 2011. This opening message is immediately, and rather unceremoniously, split into two threads with separate subjects ``parse translations" and ``lexical matters" by Makarius when he replies. It is highly unlikely, in our opinion, that an actor with a low BC measure (i.e.\  without a perceived gatekeeper power) would ever attempt to orchestrate a conversation in this way from the outset. Having said this, it is not a given whether a similar scenario would unfold if the thread had been started by a user with a much higher BC, such as Nipkow. Makarius has a BC which is around 1250 times that of Schmalz while his BC is only 3 times that of Nipkow. This aspect may need further investigation.
\begin{figure}
\begin{center}
\begin{tabular}{|@{\hskip 0.2cm}c@{\hskip 0.2cm}|@{\hskip 0.2cm}l@{\hskip 0.3cm}|@{\hskip 0.2cm}l@{\hskip 0.2cm}|}
\hline
& \isau{} & \coqc{} \\ \hline
1 &  Makarius & Chlipala \\ 
\hline
2 &  Nipkow &  Auger \\ 
\hline
3 &  Paulson &  Gross\\ 
\hline
4 &   Lochbihler & Sozeau  \\
\hline
5 &  Haftmann & Casteran \\
\hline
6 &  Huffman & Courtieu  \\
\hline
7 &  Sternagel &  Voevodsky\\
\hline
8 &  Blanchette & Spiwack \\
\hline
9 &  Krauss & Herbelin \\
\hline
10 &   Noschinski  & Vries \\
\hline
\end{tabular}
\textnormal{\caption{Betweenness centrality ranking showing the main power brokers \label{fig:bc-actors}}}
\end{center}
\end{figure}

\begin{figure}
\parbox{.45\linewidth}{
\centering
\begin{center}
\includegraphics[width=0.5\textwidth]{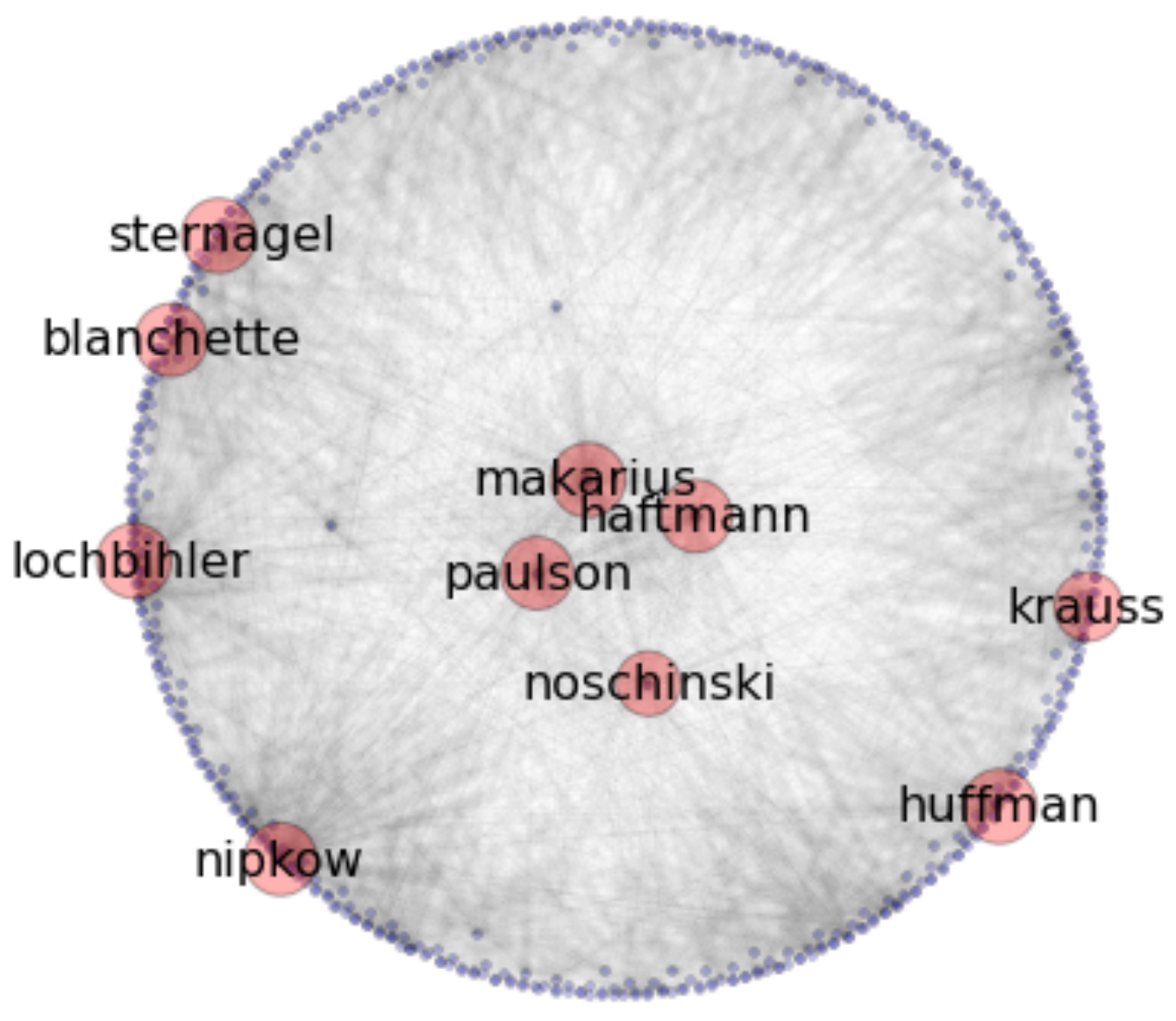}
\end{center}
\caption{Actors with BC greater than 10 times the average DC on \isau{}}
\label{fig:isa-bc-10}
}
\hfill
\parbox{.45\linewidth}{
\centering

\begin{center}
\includegraphics[width=0.5\textwidth]{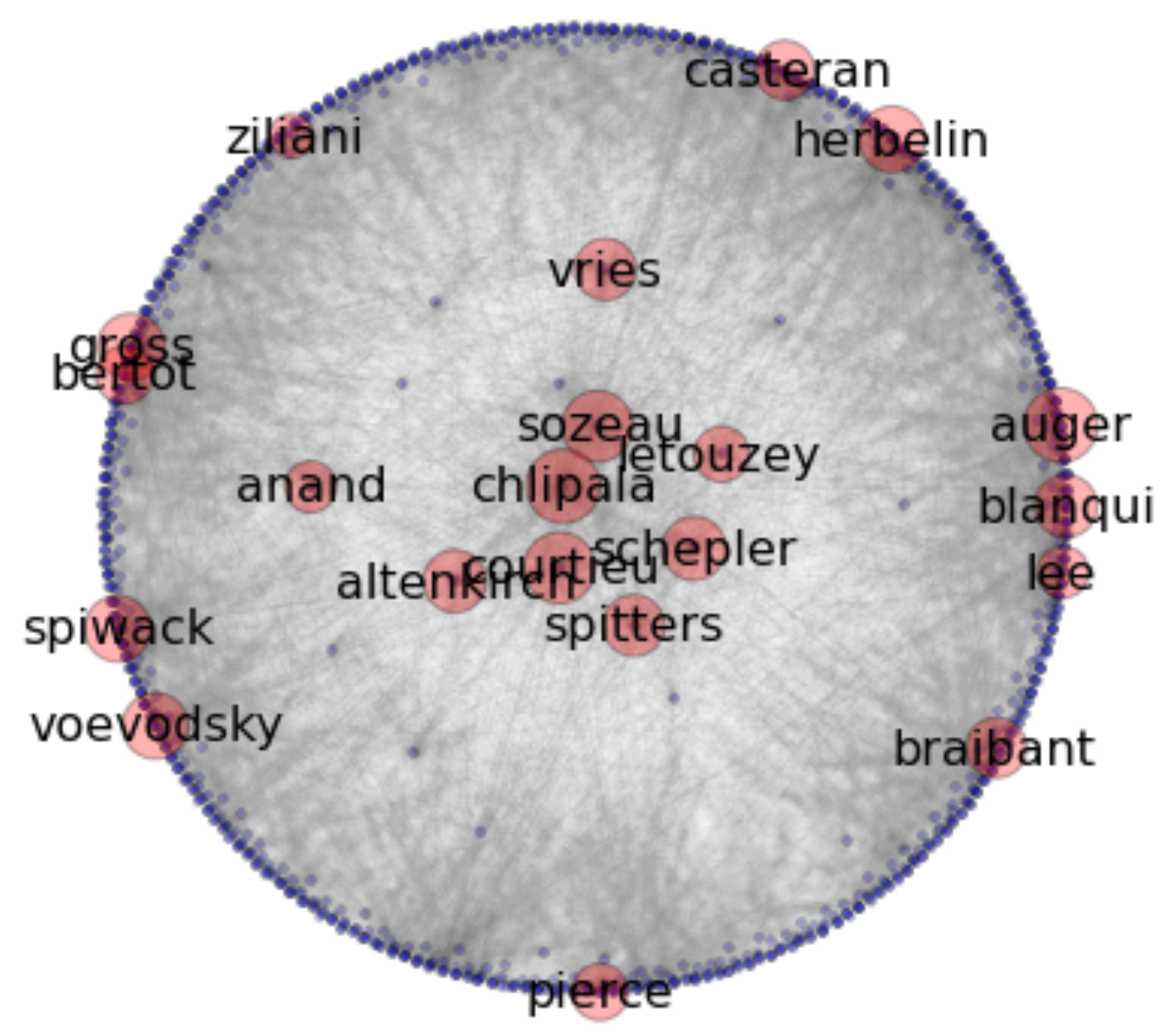}
\end{center}
\caption{Actors with BC greater than 10 times the average DC on \coqc{}}
\label{fig:coq-bc-10}

}
\end{figure}

We conclude this section with a brief but interesting remark: Figures~\ref{fig:isa-bc-10} and \ref{fig:coq-bc-10} provide a graphical overview of the number of individuals on \isau{} and \coqc{} with BC over 10 times the mean BC. It is interesting to note that, while for the Isabelle community, this basically boils down to the 10 users identified in Figure \ref{fig:bc-actors}, for the Coq community there are, at first sight, significantly more users with the potential ability to orchestrate discussions. This seems to indicate that \coqc{} is a network with more power brokers who act as focal points and influence other users' behaviours during discussions. However, when viewed as  a proportion of the total number of users in our datasets, we see that in each case this corresponds to $1.7\%$. Some further analysis revealed that at over 20 times the mean BC, the percentage was 1.0\% for both lists and at over 40 times the mean BC, it was at 0.3\%, again for both lists. This is rather surprising and may indicate that there is a general rule governing the ratio of power brokers to total number of users in such communities. 

\subsubsection{Other Measures} In the above, we have only concentrated on two measures of centrality. Aside from variations on these, there are other notions such as closeness centrality (CC), which gives an idea of how fast  an actor can reach everyone in the network, and eigenvector centrality (EC), which informally measures how well an individual is connected to other well-connected people. Although we have looked and produced interesting results and interpretations based on both of CC and EC, these will not be considered here due to space limitations. We plan to discuss these aspects  in a forthcoming paper. 
%
%

\subsection{Networking Around Topics}

In this section, we briefly look at the combination of topics extracted from messages and the network of actors that form around these. In each case, we pick a relatively interesting topic based on our knowledge of the  communities and give an overview of the individuals interested in each of them. As subject lines of messages tend to be generally well thought-out and seem sufficient \cite{bohn-r-ml11} for the task at hand, we use them to identify the relevant threads for a given topic automatically. Once we have gathered the collection of threads about the topic, we can construct the graph involving the senders and receivers who have been involved in the discussion. Note that for this approach to work, we need to assume (or check, as we did for our examples) that the identified threads generally stay on-topic, which may not always be the case for various reasons, including those discussed in Section \ref{subsubsection:bc}.


\subsubsection{Isabelle: Sledgehammer} 
\begin{figure}
\begin{center}
\includegraphics[width=0.7\textwidth]{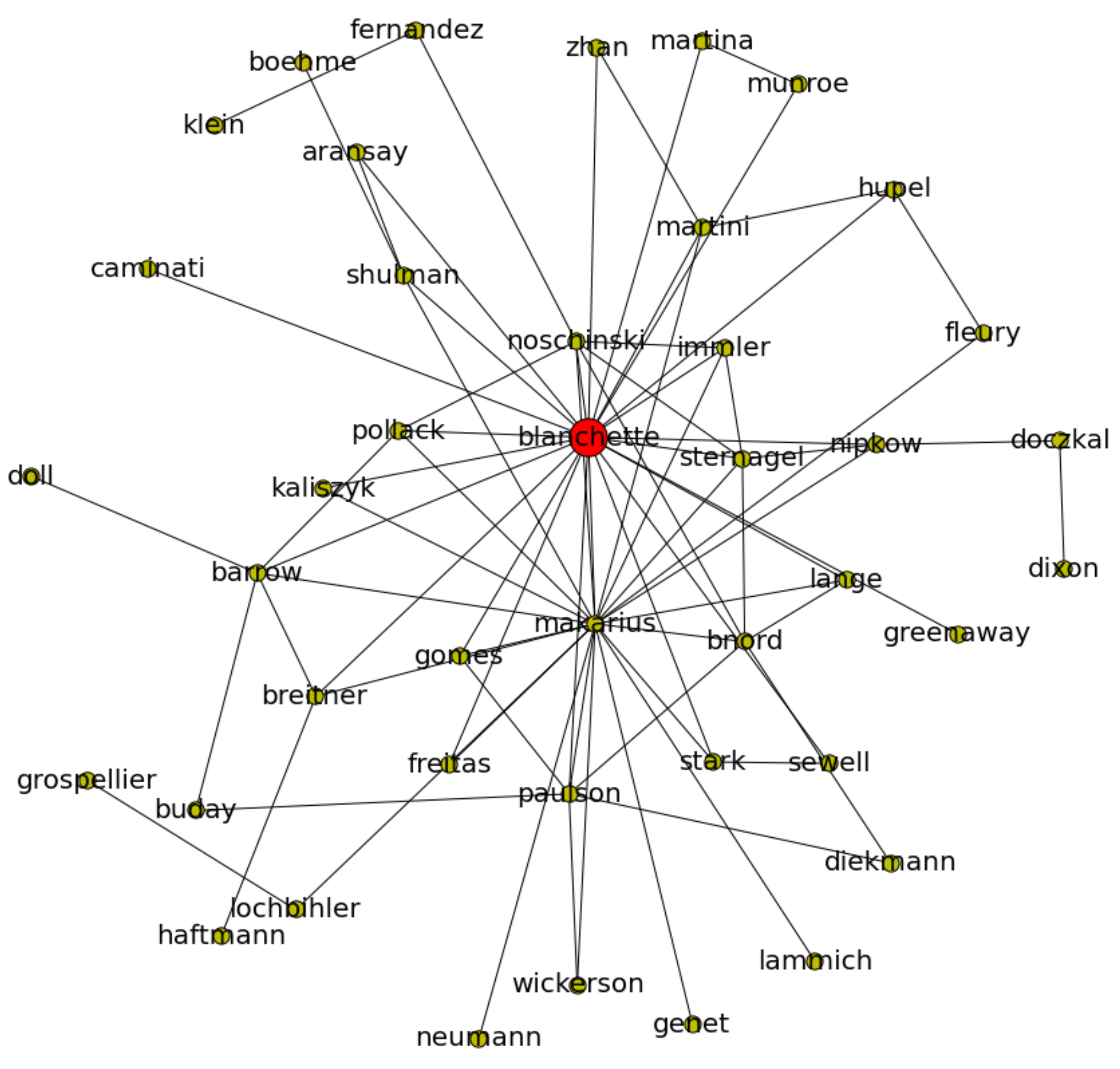}
\end{center}
\caption{The actors interested in Sledgehammer, with Blanchette as the main actor}
\label{fig:isa-sledge-egonet}
\end{figure}

Figure~\ref{fig:isa-sledge-egonet} shows the subgraph of neighbours centred around the main actor degree-wise -- Blanchette in this case --  with a radius of up to 3 hops; i.e. all neighbours and their links are included if their distance from the main actor is less than or equal to 3 when doing a breadth-first search. We use this radius as it is generally viewed as the horizon of observability for most social networks \cite{hoo-friedkin83} and a quick empirical check in our case shows no new nodes beyond 3 hops.  As one can also see, through simple inspection, Makarius is also an important hub in the sledgehammer network. An interesting question arises from this initial analysis: does this network exhibit further structure e.g.\ due to the differing expertise of the actors? We examine some of these ideas in the next section.

\subsubsection{Coq: Homotopy Type Theory (HoTT)} 
\begin{figure}
\begin{center}
\includegraphics[width=0.68\textwidth]{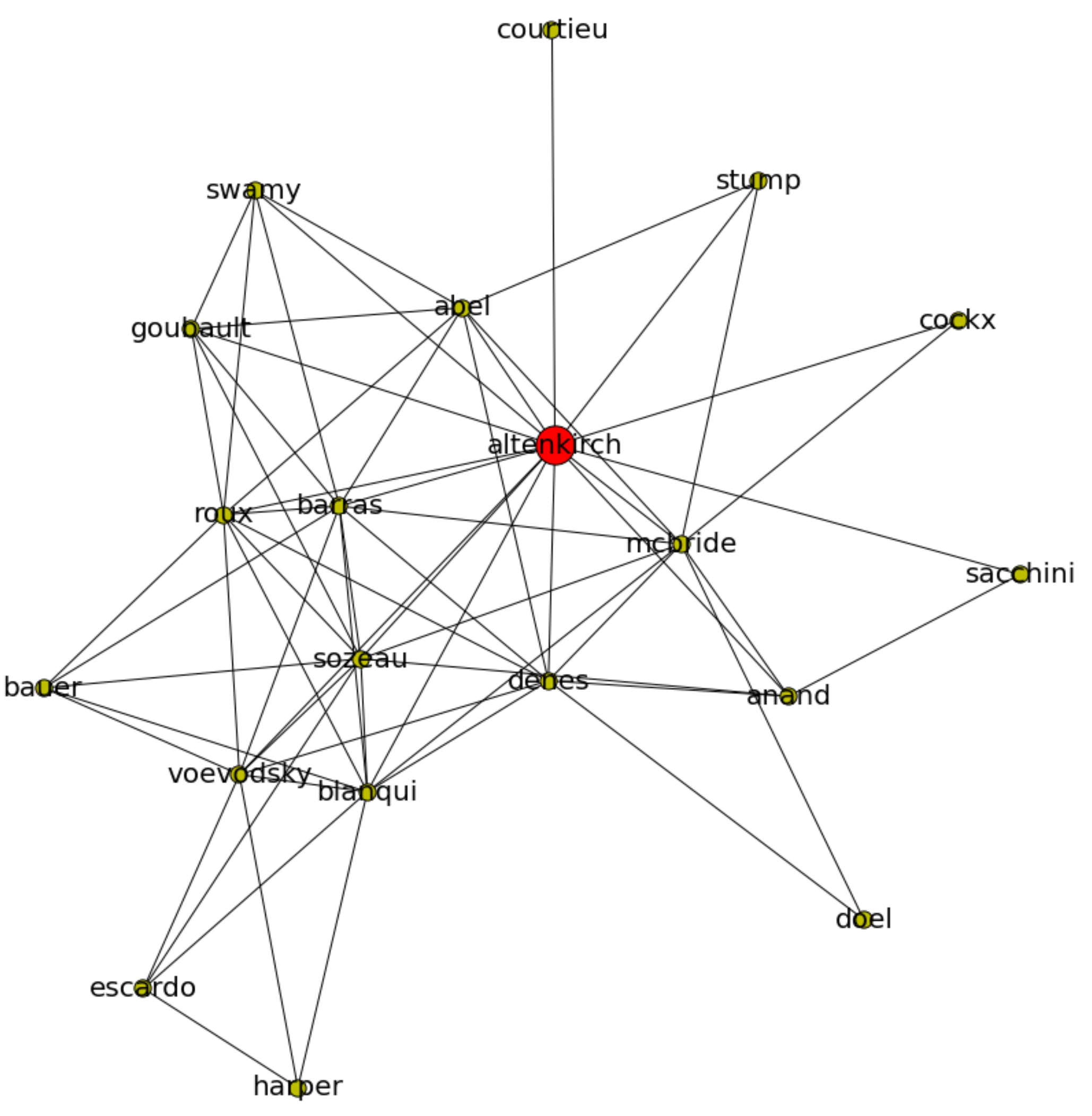}
\end{center}
\caption{The actors interested in HoTT, with Altenkirch as the main actor}
\label{fig:coq-hott-egonet}
\end{figure}

Figure~\ref{fig:coq-hott-egonet} shows the graph of main actors for messages mentioning homotopy (type) theory. In fact, our analysis shows that there is only one thread with ``homotopy'' as part of the subject lines for \coqc{} and, with a total of 86 messages, this is also the longest discussion in our Coq dataset. Again, we construct a graph based on degree-centrality, with a radius of 3. It is interesting to note that Voevodsky, though he is the undeniable reference when it comes to HoTT, is not the main actor in this discussion. One reason for this probably rests on the fact that Altenkirch is the originator of the thread and he is actually cross-posting a discussion that started on the ``Homotopy Type Theory'' mailing list (where Voevodsky is the most active poster) to \coqc{} (and to the Agda mailing list). He  even mentions the following at the end of his opening email ``I cc this to the Agda and Coq lists. Hope I haven't started a flame war''. Based on our analysis, this is a rare case of a discussion crossing over from one mailing list to another (and rather successfully in this case too), with prominent actors such as Escard{\'o} and Bauer, who are the $2^{nd}$ and $4^{th}$ most active posters on the HoTT mailing lists respectively, also joining  the conversation on \coqc. Our discussion in the next section shows that there is  more going on when we analyse the structure of this graph more closely.

\section{Discovering Subcommunities}

In this section, we give an overview of our work on exploring ``communities'' within \isau{} and \coqc{}. We look for evidence of  partitions in the global networks and delve some more into the two particular topics that we examined in the previous section to see whether the participants exhibit any  communication preferences with respect to their peers, thereby influencing the topology of the associated social graphs. 

Within the context of a graph, a community is defined as a set of nodes that are more densely connected to each other than to the rest of the network nodes; i.e. subgraphs that are densely intraconnected but sparsely inter-connected. Such community structure, also known as \emph{modularity}, is found in many real-world networks, and we have a preferred method to identify these structures in our mailing list data.

\subsection{The Louvain Method}

There are numerous algorithms (e.g.\ the well-known one developed by Clauset et al.\ \cite{PhysRevE.70.066111}) for discovering community structure in networks. In our work, we use the Louvain method \cite{blondel2008fuc}, which is a greedy algorithm that attempts to optimise the modularity of a partition of the network. In this case, modularity is defined as the fraction of edges that fall within communities minus the expected value of the same quantity if edges fall at random without regard for the community structure. The method has become popular as it often runs in $\mathcal{O}$(n log n). Informally, the algorithm  starts with a local optimisation of modularity  to find small communities. It then aggregates nodes that belong to the same community and builds a new network whose nodes are the communities. These two steps are  iterated until a maximum modularity is reached and a hierarchy of communities is produced.

\subsubsection{Isabelle: Sledgehammer Subcommunities} 
Figure~\ref{fig:hott-sub-comm} shows that the network of actors  interested in Isabelle's Sledgehammer (see Figure~\ref{fig:isa-sledge-egonet}) can also be viewed as consisting of several partitions once the Louvain method is applied. The largest one is centred around the main ego for this topic, namely Blanchette, but Makarius, Paulson and Nipkow also have their own separate subcommunities. This seems to indicate that there is some clear separation of expertise when it comes to the tool, which though originally developed by Paulson  has in more recent time been mainly under the responsibility of Blanchette. 

\begin{figure}
\begin{center}
\includegraphics[width=0.67\textwidth]{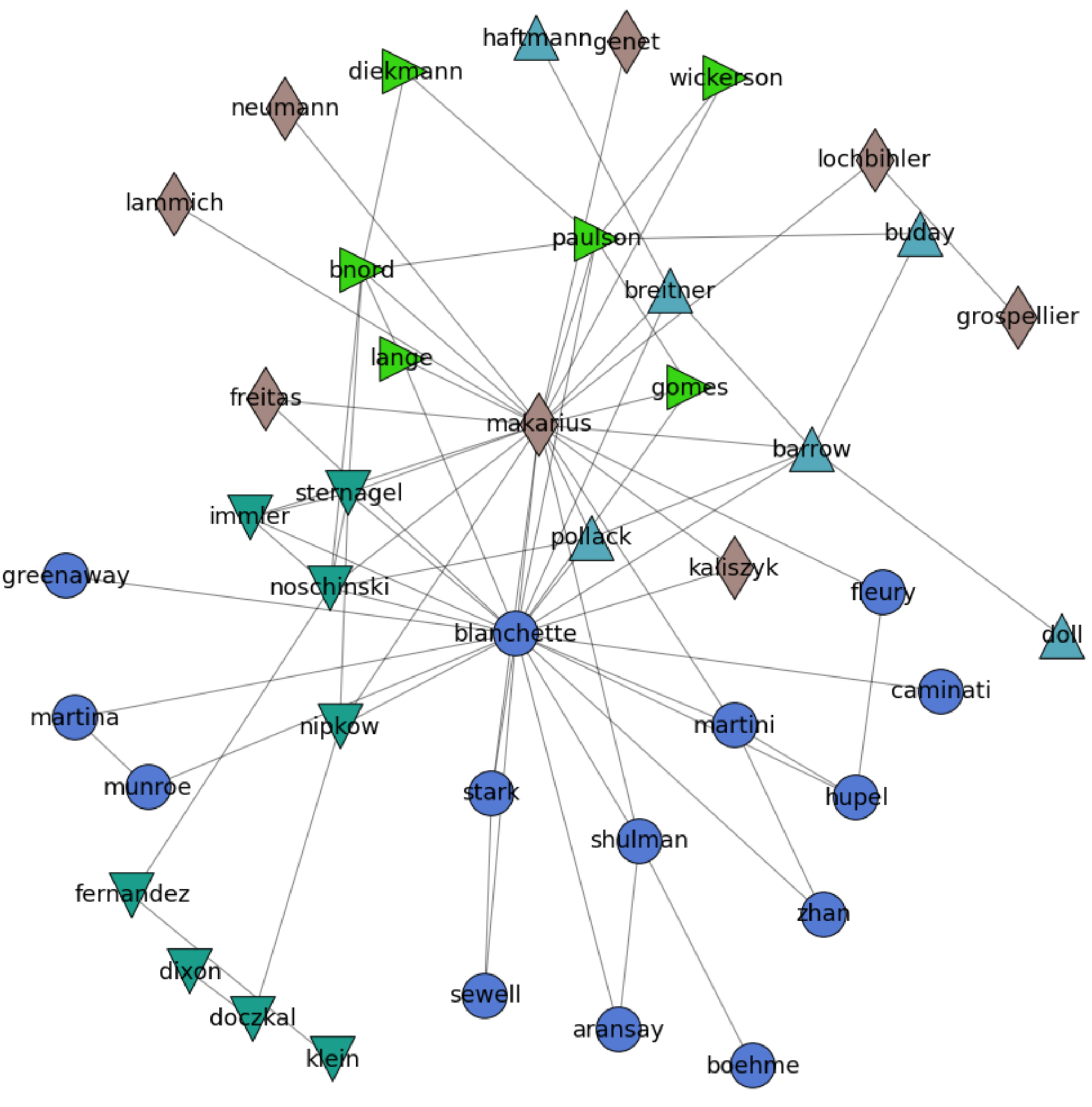}
\end{center}
\caption{The Sledgehammer network with evidence of 5 sub-communities}
\label{fig:hott-sub-comm}
\end{figure}

\subsubsection{Coq: HoTT sub-communities} Figure~\ref{fig:hott-sub-comm} shows the result of applying the community detection algorithm to the HoTT network shown in Figure~\ref{fig:coq-hott-egonet}. This  shows that the pattern of communication between the various actors is not homogeneous and that there are 3 subnetworks of users, who interact more with each other. While the one with Altenkirch, who initiated the HoTT-related thread is the largest, it is particularly interesting to see that Voevodsky, Escard{\'o} and Bauer, who already interact a lot on the HoTT mailing list, belong to the same clique of posters here. This may be an indication that their social connections have transferred over to \coqc{}. We plan to look for more occurrences of this phenomenon in other (bigger) datasets to analyse whether the inclusion of tight-knit external sub-communities can have a disruptive effect on the  network and its social processes (e.g. through an implicit foisting of views). 

\begin{figure}
\begin{center}
\includegraphics[width=0.638\textwidth]{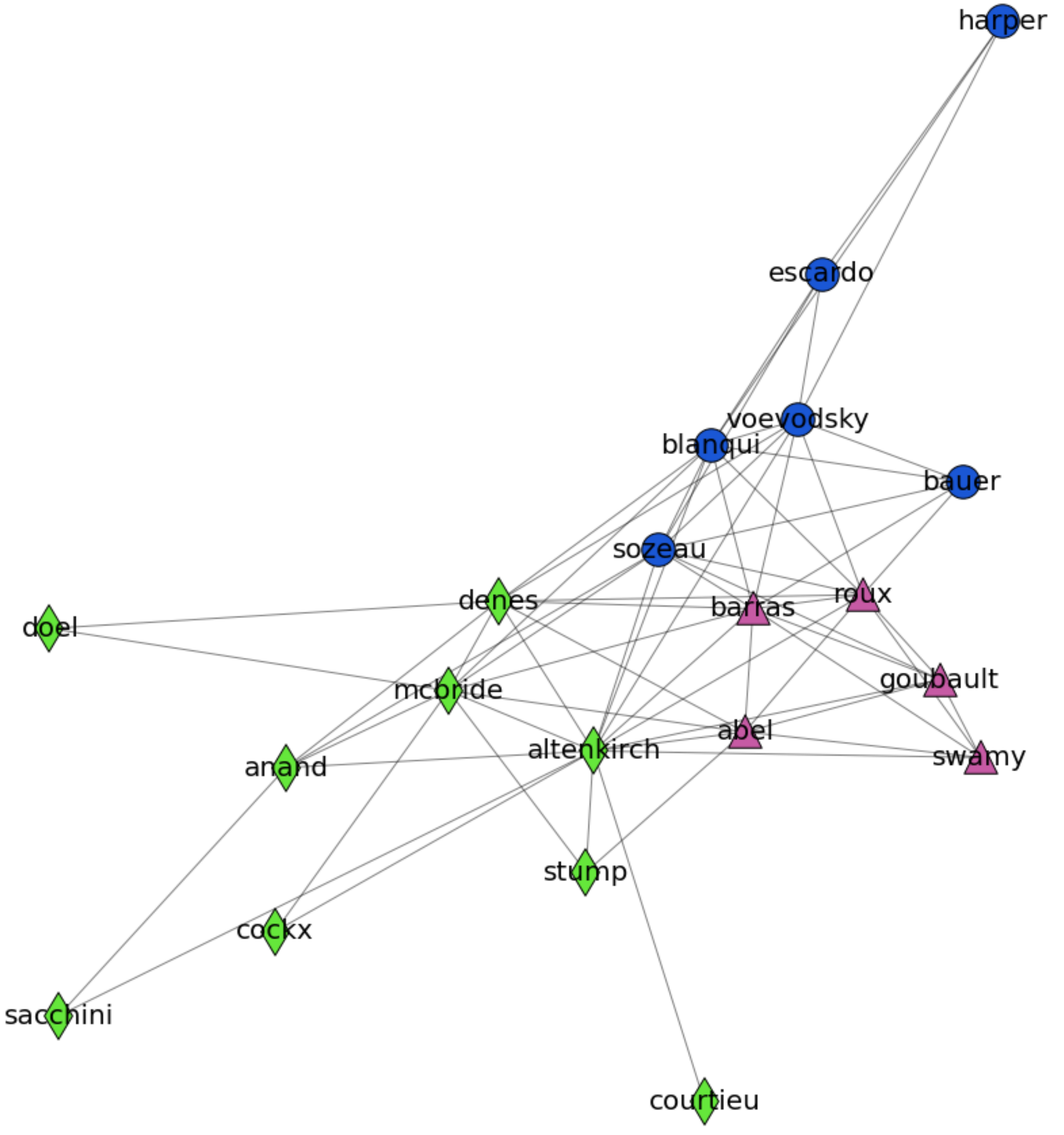}
\end{center}
\caption{The HoTT network with clear evidence of 3 sub-communities}
\label{fig:hott-sub-comm}
\end{figure}

\section{Conclusion}

We have examined various aspects of the structure of the interactive theorem proving communities as reflected in the \isau{} and 
\coqc{} mailing lists. The structures we discovered seem plausible and overlap well enough with our a-priori expectations of how they would look like. There are many more insights to be discovered from our body of data. For instance, we have worked on the automatic discovery of topics, which is not described in this paper because of space constraints. We are also working on building a fine-grained understanding of how the \isau{} and \coqc{} networks evolve over time. Their users can join and leave at any time and, although not explicitly addressed in this paper, it seems clear that some of the above-mentioned actors have not been as active recently as they used to be (e.g.\ after obtaining their PhD or moving to a different job), making their current influence in their networks less clear in some cases.

Our focus for forthcoming work will be how to apply the insights gained from this work to building the social engine for ProofPeer, our system in the making for collaborative theorem proving~\cite{proofpeerpositionpaper}. Immediately applicable are our experiences from extracting our graph models from the raw data and applying a range of algorithms. In order to scale, we need this process to be fully automatic in ProofPeer, and without incurring the large amount of manual labour necessary for this paper. Therefore we are designing the communication mechanisms in ProofPeer in a way which allows us to easily and automatically obtain the data we need. 

The more difficult step is to integrate the insights gained from our study with the reputation and recommender systems we are developing for ProofPeer. We plan to measure reputation in ProofPeer by a multidimensional vector, with one or more components of this vector influenced by appropriately weighted notions from social network analysis such as betweenness centrality of informal communication and reciprocity.
Automatic recommendation is then based in part on reputation but also on notions such as induced sub-communities: artefacts from reputable and pertinent sources are more likely to be recommended than those from less reputable or loosely-related ones.

\section*{Acknowledgement}
This work is supported by EPSRC grant EP/L011794/1.

\bibliography{proofpeer-sna}{}
\bibliographystyle{plain}

\end{document}